\documentclass[runningheads]{llncs}

\usepackage[T1]{fontenc}

\usepackage{graphicx}
\usepackage{hyperref}
\usepackage{amsmath}
\usepackage{longtable}
\usepackage{booktabs} 
\usepackage{xcolor}
\usepackage{float}
\usepackage{dashrule}
\usepackage{ulem}
\usepackage{dsfont}
\usepackage{graphicx}
\usepackage{caption}

\DeclareMathOperator*{\argmax}{arg\,max}

\definecolor{violet}{RGB}{148, 0, 211}

\definecolor{purple}{RGB}{128, 0, 128}

\newcommand{\commentout}[1]{}

\begin{document}
\title{Inducing Diversity in Differentiable Search Indexing}

\newcommand{\repeatthanks}{\textsuperscript{\thefootnote}}
\author{Abhijeet Phatak\inst{1}\orcidID{0000-0002-6856-4799}\thanks{These authors contributed equally to this research.} \and
Jayant Sachdev\inst{1}\orcidID{0009-0000-1834-1338}\repeatthanks \and
Sean D Rosario\inst{1}\orcidID{0009-0003-5033-9395}\repeatthanks \and
Swati Kirti\inst{1}\orcidID{0009-0008-4424-4695} \and
Chittaranjan Tripathy\inst{1}\orcidID{0009-0006-7353-4793}
}

\authorrunning{A. Phatak et al.}

\institute{\href{https://tech.walmart.com/content/walmart-global-tech/en_us.html}{Walmart Global Tech}\\
640 W California Avenue, Sunnyvale, CA, USA 94086 \newline 
Corresponding Author : \email{abhijeet.phatak@walmart.com}
}

\maketitle              
\begin{abstract}
Differentiable Search Indexing (DSI) is a recent paradigm for information retrieval which uses a transformer-based neural network architecture as the document index to simplify the retrieval process. A differentiable index has many advantages enabling modifications, updates or extensions to the index. In this work, we explore balancing relevance and novel information content (diversity) for training DSI systems inspired by Maximal Marginal Relevance (MMR), and show the benefits of our approach over the naive DSI training. We present quantitative and qualitative evaluations of relevance and diversity measures obtained using our method on NQ320K and MSMARCO datasets in comparison to naive DSI. With our approach, it is possible to achieve diversity without any significant impact to relevance. Since we induce diversity while training DSI, the trained model has learned to diversify while being relevant. This obviates the need for a post-processing step to induce diversity in the recall set as typically performed using MMR. Our approach will be useful for Information Retrieval problems where both relevance and diversity are important such as in sub-topic retrieval. Our work can also be easily be extended to the incremental DSI settings which would enable fast updates to the index while retrieving a diverse recall set.
\end{abstract}

\section{Introduction}
Recent developments in Information Retrieval (IR) have brought forth the Differentiable Search Index (DSI) \cite{tay2022transformer}, marking a departure from traditional retrieval methodologies. Conventional approaches typically involve constructing an external index to facilitate the retrieval process. In contrast, DSI maps user queries directly to relevant documents with a neural network, streamlining the retrieval pipeline. This innovation offers significant benefits including ease of implementation, reduced index storage requirements, and the capability for end-to-end retrieval. It is possible to incrementally update such an index in a short time with new documents without requiring a complete re-indexing which could be time consuming \cite{kishore2023incdsi,dsi++}. However, despite its groundbreaking retrieval mechanism, current DSI models still rely on relevance signals from query-document pairs during training. Unlike dense retrieval models that enable explicit interactions between queries and documents at retrieval time, DSI models map short textual inputs directly to specific document identifiers without such interactions. Moreover, lexical and semantic overlap among the retrieved documents can result in a lack of diversity in the final outputs. To address these limitations, this study aims to introduce methods for enhancing diversity in the set of retrieved documents. Below are our \textbf{key contributions} in this paper:\vspace{-0.05in}
\begin{itemize}
\item We introduce an extension to Differentiable Search Index (DSI) that enables the model to diversify the retrieved documents.
\item By introducing a novel loss function which adds a diversity component inspired by document similarity from Maximal Marginal Relevance (MMR) that induces diversity during model training.
\item We show that our approach can retrieve documents that are both relevant and diverse, with no additional latency during inference nor requiring additional post-processing steps.
\end{itemize}

\section{Related Work}
\vspace{-0.10in}
While traditional methods for diversification are mainly manually crafted, recent research in this area shifts to supervised learning methods and shows superior performance on diversity evaluation metrics. Most current approaches employ an iterative approach, where the next document is selected among remaining documents to maximize some objective with respect to the ones already selected \cite{ResDivSearchReco}. Such a paradigm is intuitive and achieves the task of diversification. However, the main challenge is that learning is inherently less effective because there is an exponentially large number of ranking lists to consider.
Previous approaches such as \cite{R-LTR} focus on the ideal diversified ranking lists. Reinforcement learning (RL) based approaches such as \cite{divRankPolicyNets} and \cite{mdpSearchDiversify} try to maximize the expected rewards over sampled lists from a distribution. Recently proposed PAMM \cite{MMRSearchRelevanceDivMeasures} and DVGAN \cite{liu2020dvgan} methods maximize the margin between sampled positive and negative lists during training and show better performance. However, gathering high-quality samples is challenging due the scale of data. Prior work incorporates diversity loss into Learning-To-Rank training by approximating diversity metrics \cite{DALeToR}.

\section{Background}
\subsection{Differentiable Search Index}
Differentiable search indexing (DSI) \cite{tay2022transformer} is a paradigm for information retrieval with transformer neural network models, where information about a collection of documents is encoded in the model parameters. There is no separate index apart from the neural network in the DSI paradigm. Given a query, the model predicts the documents ID(s) of the relevant document(s), using a classifier like a softmax over document ids or a decoder that generates document ID(s).

\subsection{Maximal Marginal Relevance}
Maximum Marginal Relevance (MMR) \cite{mmr} retrieval iteratively finds documents that are dissimilar to previous results. MMR has been shown to improve retrievals in search, recommendations and LLMs \cite{mmrsearch,mmrrecs,mmrinllm}. Consider the set $D$ consisting of all candidate documents and $R$ consisting of the previously chosen (retrieved) documents, and $q$ representing a query. We also consider two similarity functions -- $Sim_1$ that compares a document with the query, and $Sim_2$ that assesses the similarity between two documents. Let $d_i$ and $d_j$ denote individual documents in $D$ and $R$.  Then MMR is defined as 
\begin{eqnarray*}
\text{{MMR}} &=& \argmax_{d_i \in D \setminus R}  \lambda_{MMR} \cdot Sim_1(d_i, q) - (1 - \lambda_{MMR}) \cdot \max_{d_j \in R} Sim_2(d_i, d_j),
\end{eqnarray*}
where the first term and the second term respectively represent relevance and diversity. The adjustable parameter $\lambda_{MMR}$ balances the importance of relevance and diversity. $\lambda_{MMR} \rightarrow 1$ prioritizes relevance whereas $\lambda_{MMR} \rightarrow 0$ prioritizes diversity.

\subsection{Diversity in Information Retrieval}
Diversity is important in information retrieval systems like search engines and recommendation systems. With corpora containing large number of documents, there is a high likelihood that the retrieved set of documents would contain duplicate documents or partially duplicate documents that contain a high degree of textual or semantic overlap.
Without diversity, users might get an incomplete picture of a topic or biased understanding. Diversity in the retrieved set could cater to a wider range of interests and need, while mitigating risks of over-specialization/bias and ensures a larger coverage. With sub-topic retrieval, it also ensures that all relevant subtopics are adequately represented, providing users with a more well-rounded understanding of the main topic. Thus, diversity is a key component in making information retrieval more effective and user friendly.

\section{Methods and Experiments}
\subsection{Datasets}
In our experiments, we adopt the same data and settings outlined in the IncDSI paper, utilizing two publicly available datasets. The Natural Questions 320K (NQ320K) dataset \cite{kwiatkowski2019natural} contains pairs of queries and documents, where queries are natural language questions and the documents are Wikipedia articles that provide the corresponding answers. MSMARCO Document Ranking \cite{nguyen2016ms}, another prominent dataset for question-answering tasks, consists of search queries and 3.2 million documents, but only a subset of these documents are directly associated with queries. In both datasets, each document is assigned a distinct document identifier (docid). We use the data split given by the IncDSI authors, but only train and evaluate on the data they use for the base DSI model.
\vspace{-0.1in}
\subsection{Our approach: Inducing diversity in DSI}
Our approach to inducing diversity in the model is achieved by modifying the loss function in the training step. The naive DSI model used as our baseline is the BERT-based classification model outlined in IncDSI\cite{kishore2023incdsi}, which uses a multi-class cross entropy loss for classification, where the classes are the $N$ document labels and $p_i$ is predicted probability and $y_i$ is the ground truth label. We extend the loss with the second component inspired from MMR, which accounts for similarity within the retrieved set of documents. In our implementation, we select the linear layer for the top $K$ logits predicted -- which represent the top scoring documents for a query, and then calculate averaged self-similarity within that set of document representations using cosine similarity. The total loss is a linear combination of the two components, weighted by a factor of $\alpha$ and is defined as
\begin{eqnarray*}
 Loss_{total} &=& \alpha \cdot (-\frac{1}{N} \sum_{i=1}^{N} y_i \log(p_i)) + (1 - \alpha) \cdot \sum_{d_i,d_j \in K, j>i} Sim_2(d_i, d_j).
\end{eqnarray*}
When $\alpha=1$, the diversity component is not included in the loss function and thus it corresponds to the naive DSI setting.

\vspace{-0.25in}
\begin{figure}[h]
\centering
\includegraphics[width=3.9in]{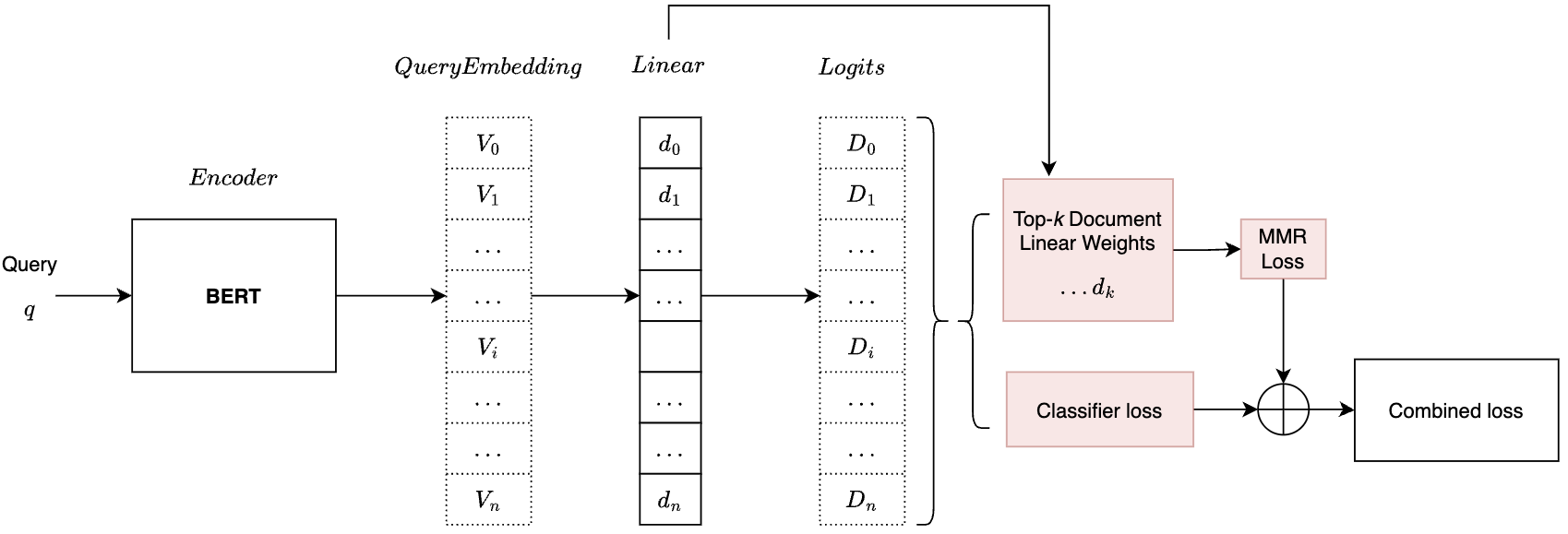}
\caption{Overview of our proposed approach to induce diversity in DSI}
\end{figure}

\vspace{-0.4in}
\subsection{Experimental Setup}
We use the BERT model \cite{Devlin2019BERTPO} and initialized with publicly available bert-base-uncased weights for all our experiments. The classification layer is randomly initialized. For the DSI baseline, we use the classifier-based DSI implementation from the IncDSI paper. We train 20 epochs for all models, and train model with $\alpha \in \{0.25, 0.5, 0.75, 1\}$. We use the same learning rate of $1e^{-5}$ and $5e^{-5}$ and a batch size of 128 and 512 are used for NQ320K and MSMARCO respectively mentioned in the IncDSI paper. We use a sigle 80GB CUDA compatible GPU for training and validation of all experiments.

\subsection{Metrics}
We evaluate several relevance and diversity metrics as described in Table ~\ref{table:metrics}.
\begin{table}[H]
\centering
\scalebox{0.88}{
\begin{tabular}
{|p{0.14\linewidth}|p{0.38\linewidth}|p{0.6\linewidth}|}
\hline
\textbf{Metric}  &  \textbf{Formula}  & \textbf{Notes}\\
\hline
Hits@$K$  & $Hits@k = \frac{1}{Q} \sum_{i=1}^{Q} \mathds{1}(\text{rank}_i \leq k)$  & Accuracy of top-$K$ ($K \in \{1,5,10\}$) items of retrieved set. Higher Hits@$K$ means that the model is more effective at retrieving relevant results.\\
\hline
\addlinespace[0.2em] 
MRR@$K$  & $\text{MRR@k} = \frac{1}{Q} \sum_{i=1}^{Q} \frac{1}{rank_i}$  & Mean Reciprocal Rank at $K(=10)$ is a statistical measure for evaluating documents retrieved for queries, ordered by probability of correctness. The reciprocal rank of documents for a query is the inverse of the rank of the first correct answer.\\
\hline
\hline
\addlinespace[0.2em]
ROUGE-L ($\text{hom}(\{D\})$) & $\frac{1}{|D| - 1}\ \sum_{d, d' \in \{D\}; d \neq d'}\text{sim}(d, d')$  & Homogenization score calculates overlap between outputs by identifying the longest common subsequence of words. A high ROUGE-L score suggests that the outputs have significant structural repetition, implying low diversity. $\text{sim}(d, d')$ represents the ROUGE-L similarity between two different documents $d$ and $d'$ computed using uncased DistilBERT embeddings.\\
\hline
\addlinespace[0.2em]
${\text{NGD}}(\{D\})$ &  $\sum_{n=1}^{4} \frac{\text{\# unique n-grams in } \{D\}}{\text{\# total n-grams in } \{D\}}$ &  N-Gram Diversity score extends the idea of lexical diversity by evaluating sequences of words (n-grams) rather than individual words. It computes the ratio of unique n-grams (up to length four) to the total number of n-grams in the text, with higher NGD values indicating less repetitive sequences.\\
\hline
\addlinespace[0.2em]
$\text{CR}(\{D\})$  &  $\frac{\text{size of } \{D\}}{\text{compressed size of } \{D\}}$ & Compression Ratio (CR) evaluate how well a text can be compressed using algorithms like gZip. Text that compresses more effectively is likely to contain more redundancy and repetition. A higher compression ratio reflects lower diversity. The numerator represents the original size of the set $\{D\}$, and the denominator represents the size after compression. A compression ratio close to 1 indicates maximal diversity.\\
\hline
\end{tabular}
}
\caption{Relevance and Diversity metrics.}
\label{table:metrics}
\end{table}

\vspace{-.5in}
\section{Results and Discussion}
\vspace{-0.05in}
We observe that our method performed similar to naive DSI in terms of Hits and MRR. However, the compression ratios and other diversity metrics indicate that in our setting with induced diversity in DSI (where $\alpha<1$) we retrieve more diverse documents related to the query with similar performance on relevance metrics. In most cases, the diversity-induced models slightly outperform the naive DSI model. The mean inference time across all models is 0.65 seconds with a standard deviation of 0.017 seconds. Since the same number of model parameters are used in both the cases, with and without diversity, the inference time is not impacted with our method.

\begin{table}[htbp]
\centering
\begin{tabular}
{|@{\hspace{2pt}}l@{\hspace{2pt}}|@{\hspace{2pt}}c@{\hspace{2pt}}|@{\hspace{2pt}}c@{\hspace{2pt}}|@{\hspace{2pt}}c@{\hspace{2pt}}|@{\hspace{2pt}}c@{\hspace{2pt}}|@{\hspace{2pt}}c@{\hspace{2pt}}|@{\hspace{2pt}}c@{\hspace{2pt}}|@{\hspace{2pt}}c@{\hspace{2pt}}|@{\hspace{2pt}}c@{\hspace{2pt}}|}
\hline
\textbf{Dataset}  & \textbf{$\alpha$}  & \textbf{Hits@1}  & \textbf{Hits@5}  & \textbf{Hits@10}  & \textbf{MRR@10}  & \textbf{ROUGE-L}  & \textbf{NGD}  & \textbf{CR}\\
\hline
MSMARCO  & 1  & 0.4837  & 0.7419  & 0.8093 & 0.5937  & 0.165  & 4.509  & 1.204 \\
MSMARCO  & 0.75  & 0.4664  & 0.7240  & 0.7969  & 0.5757  & 0.088  & 4.750  & 1.242\\
MSMARCO  & 0.50  & 0.4856  & 0.7415  & 0.8093  & 0.5949  & 0.148  & 4.528  & 1.082 \\
MSMARCO  & 0.25  & 0.4826  & 0.7430  & 0.8123  & 0.5929  & 0.188  & 4.646  & 1.093 \\
\hline
NQ320K  & 1 & 0.6747  & 0.8445  & 0.8749  & 0.7486 & 0.171 & 3.834  & 2.552 \\
NQ320K  & 0.75  & 0.6753  & 0.8496  & 0.8812  & 0.7518  & 0.138  & 4.167  & 2.323 \\
NQ320K  & 0.50  & 0.6769  & 0.8486  & 0.8828  & 0.7526  & 0.132  & 4.193  & 2.266 \\
NQ320K  & 0.25  & 0.6763  & 0.8492  & 0.8808  & 0.7518  & 0.138  & 4.165  & 2.333 \\
\hline
\end{tabular}
\caption{Results on NQ320K and MSMARCO.}
\label{table:results}
\vspace{-0.35in}
\end{table}

We also did a qualitative analysis to ensure that the documents retrieved by the new model with diversity is more diverse than DSI-scratch. Consider the generic query ``California''. Table \ref{table:examples} shows the difference in retrieved documents between the two models trained on NQ320K dataset. While both sets cover the history, demographics, and economy of California, Document set 2 also includes information about specific regions such as the Monterey Peninsula and Sacramento, and even a popular song about California. This variety of topics makes Document set 2 more diverse.

\begin{table}[htbp]
  \centering
   \captionsetup{skip=2pt}
  \scalebox{0.85}{%
    \begin{tabular}{|@{\hspace{5pt}}l@{\hspace{5pt}}|@{\hspace{5pt}}l@{\hspace{5pt}}|}
    \hline
    \textbf{Document set 1 ($\alpha=1$)} & \textbf{Document set 2 ($\alpha=0.5$)} \\
    \hline
     1. Hispanics and Latinos in California &  1. Hispanics and Latinos in California \\
    2. List of people from California & 2. List of people from California \\
    3. Californio & 3. Calafia \\
    4. California & 4. List of cities and towns in California \\
    5. List of cities and towns in California & 5. Demographics of California \\
    6. History of California & 6. Economy of California \\
    7. Sports in California & 7. California \\
    8. Economy of California & 8. Monterey Peninsula \\
    9. History of Los Angeles & 9. Sacramento, California \\
    10. Fauna of California & 10. Going Back to Cali (The Notorious B.I.G. song) \\
    \hline
    \end{tabular}%
  }
\caption{Examples of documents retrieved for the query ``California'' without ($\alpha=1$) and with ($\alpha=0.5$) induced diversity}
\vspace{-0.65in}
\label{table:examples}
\end{table}

\section{Conclusion}
Our work takes the Differentiable Search Index (DSI) framework for information retrieval, and adds a component to induce diversity in the search results. Since the diversity component is included in the training step, this obviates additional computational steps post retrieval such as MMR, beam search or other mechanisms used to diversify retrieved results. Our work can be extended to IncDSI (incremental) setting where the index can be updated with new document streams, and is an interest direction to pursue. We run experiments on the NQ320K and MSMARCO datasets, with varying values of the balance of relevance and diversity varied by the parameter $\alpha$. We demonstrate that our approach can induce diversity in the search results without significant impact on the relevance.

\bibliographystyle{splncs04}
\bibliography{mybibliography}

\begin{thebibliography}{10}
\providecommand{\url}[1]{\texttt{#1}}
\providecommand{\urlprefix}{URL }
\providecommand{\doi}[1]{https://doi.org/#1}

\bibitem{mmr}
Carbonell, J., Goldstein, J.: The use of mmr, diversity-based reranking for reordering documents and producing summaries. In: Proceedings of the 21st annual international ACM SIGIR conference on Research and development in information retrieval. pp. 335--336 (1998)

\bibitem{Devlin2019BERTPO}
Devlin, J., Chang, M.W., Lee, K., Toutanova, K.: Bert: Pre-training of deep bidirectional transformers for language understanding. In: North American Chapter of the Association for Computational Linguistics (2019), \url{https://api.semanticscholar.org/CorpusID:52967399}

\bibitem{divRankPolicyNets}
Feng, Y., Xu, J., Lan, Y., Guo, J., Zeng, W., Cheng, X.: From greedy selection to exploratory decision-making: Diverse ranking with policy-value networks. In: The 41st International ACM SIGIR Conference on Research \& Development in Information Retrieval. p. 125–134. SIGIR '18, Association for Computing Machinery, New York, NY, USA (2018). \doi{10.1145/3209978.3209979}, \url{https://doi.org/10.1145/3209978.3209979}

\bibitem{mmrsearch}
Guo, S., Sanner, S.: Probabilistic latent maximal marginal relevance. In: Proceedings of the 33rd international ACM SIGIR conference on Research and development in information retrieval. pp. 833--834 (2010)

\bibitem{kishore2023incdsi}
Kishore, V., Wan, C., Lovelace, J., Artzi, Y., Weinberger, K.Q.: Incdsi: incrementally updatable document retrieval. In: International Conference on Machine Learning. pp. 17122--17134. PMLR (2023)

\bibitem{kwiatkowski2019natural}
Kwiatkowski, T., Palomaki, J., Redfield, O., Collins, M., Parikh, A., Alberti, C., Epstein, D., Polosukhin, I., Devlin, J., Lee, K., et~al.: Natural questions: a benchmark for question answering research. Transactions of the Association for Computational Linguistics  \textbf{7},  453--466 (2019)

\bibitem{liu2020dvgan}
Liu, J., Dou, Z., Wang, X., Lu, S., Wen, J.R.: Dvgan: A minimax game for search result diversification combining explicit and implicit features. In: Proceedings of the 43rd International ACM SIGIR Conference on Research and Development in Information Retrieval. pp. 479--488 (2020)

\bibitem{dsi++}
Mehta, S.V., Gupta, J., Tay, Y., Dehghani, M., Tran, V.Q., Rao, J., Najork, M., Strubell, E., Metzler, D.: Dsi++: Updating transformer memory with new documents. arXiv preprint arXiv:2212.09744  (2022)

\bibitem{nguyen2016ms}
Nguyen, T.: Ms marco: A human generated machine reading comprehension dataset. arXiv preprint arXiv:1611.09268  (2016)

\bibitem{tay2022transformer}
Tay, Y., Tran, V., Dehghani, M., Ni, J., Bahri, D., Mehta, H., Qin, Z., Hui, K., Zhao, Z., Gupta, J., et~al.: Transformer memory as a differentiable search index. Advances in Neural Information Processing Systems  \textbf{35},  21831--21843 (2022)

\bibitem{mmrrecs}
Wu, C.H., Wang, Y., Ma, J.: Maximal marginal relevance-based recommendation for product customisation. Enterprise Information Systems  \textbf{17}(5),  1992018 (2023)

\bibitem{ResDivSearchReco}
Wu, H., Zhang, Y., Ma, C., Lyu, F., He, B., Mitra, B., Liu, X.: Result diversification in search and recommendation: A survey. IEEE Trans. on Knowl. and Data Eng.  \textbf{36}(10),  5354–5373 (Apr 2024). \doi{10.1109/TKDE.2024.3382262}, \url{https://doi.org/10.1109/TKDE.2024.3382262}

\bibitem{MMRSearchRelevanceDivMeasures}
Xia, L., Xu, J., Lan, Y., Guo, J., Cheng, X.: Learning maximal marginal relevance model via directly optimizing diversity evaluation measures. In: Proceedings of the 38th international ACM SIGIR conference on research and development in information retrieval. pp. 113--122 (2015)

\bibitem{mdpSearchDiversify}
Xia, L., Xu, J., Lan, Y., Guo, J., Zeng, W., Cheng, X.: Adapting markov decision process for search result diversification. In: Proceedings of the 40th International ACM SIGIR Conference on Research and Development in Information Retrieval. p. 535–544. SIGIR '17, Association for Computing Machinery, New York, NY, USA (2017). \doi{10.1145/3077136.3080775}, \url{https://doi.org/10.1145/3077136.3080775}

\bibitem{DALeToR}
Yan, L., Qin, Z., Pasumarthi, R.K., Wang, X., Bendersky, M.: Diversification-aware learning to rank using distributed representation. In: Proceedings of the Web Conference 2021. p. 127–136. WWW '21, Association for Computing Machinery, New York, NY, USA (2021). \doi{10.1145/3442381.3449831}, \url{https://doi.org/10.1145/3442381.3449831}

\bibitem{mmrinllm}
Ye, X., Iyer, S., Celikyilmaz, A., Stoyanov, V., Durrett, G., Pasunuru, R.: Complementary explanations for effective in-context learning. arXiv preprint arXiv:2211.13892  (2022)

\bibitem{R-LTR}
Zhu, Y., Lan, Y., Guo, J., Cheng, X., Niu, S.: Learning for search result diversification. In: Proceedings of the 37th International ACM SIGIR Conference on Research \& Development in Information Retrieval. p. 293–302. SIGIR '14, Association for Computing Machinery, New York, NY, USA (2014). \doi{10.1145/2600428.2609634}, \url{https://doi.org/10.1145/2600428.2609634}

\end{thebibliography}


\end{document}